\documentclass[journal,12pt,draftcls,onecolumn]{IEEEtran}

\usepackage{cite}
\ifCLASSINFOpdf
  \usepackage[pdftex]{graphicx}
\fi
\usepackage[cmex10]{amsmath}
\usepackage{array}
\usepackage{mdwmath}
\usepackage{mdwtab}
\usepackage{url}
\usepackage[caption=false,font=footnotesize]{subfig}
\usepackage{multirow}

\begin{document}
\title{Analysis of snowpack properties and structure from TerraSAR-X data, based on multilayer backscattering and snow evolution modeling approaches}

\author{Xuan-Vu~Phan,
        Laurent~Ferro-Famil,~\IEEEmembership{Member,~IEEE,}
        Michel~Gay,~\IEEEmembership{Member,~IEEE,}
        Yves~Durand, 
	Marie Dumont, 
	Sophie~Allain
	and~Guy~D'Urso% <-this % stops a space
\thanks{Xuan-Vu Phan and Michel Gay are with the Grenoble Image Parole Signal et Automatique lab, Grenoble, France.}% <-this % stops a space
\thanks{Laurent Ferro-Famil and Sophie Allain are with the Institut d'Electronique et de T\'el\'ecommunications de Rennes, University of Rennes, France.}% <-this % stops a space
\thanks{Yves Durand and Marie Dumont is with M\'et\'eo-France and CNRS, CNRM-GAME, URA-1357, Centre d'Etude de la Neige, France.}% <-this % stops a space
\thanks{Guy D'Urso is with Electricit\'e de France, Paris, France.}
}
\maketitle

\begin{abstract}
%\boldmath
Recently launched high precision Synthetic Aperture Radar (SAR) satellites such as TerraSAR-X, COSMO-SkyMed, etc. present a high potential for better observation and characterization of the cryosphere. This study introduces a new approach using high frequency (X-band) SAR data and an Electromagnetic Backscattering Model (EBM) to constrain the detailed snowpack model Crocus. A snowpack EBM based on radiative transfer theory, previously used for C-band applications, is adapted for the X-band. From measured or simulated snowpack stratigraphic profiles consisting of snow optical grain radius and density, this forward model calculates the backscattering coefficient $\sigma^0$ for different polarimetric channels. The output result is then compared with spaceborne TerraSAR-X acquisitions to evaluate the forward model. Next, from the EBM, the adjoint operator is developed and used in a variational analysis scheme in order to minimize the discrepancies between simulations and SAR observations. A time series of TerraSAR-X acquisitions and in-situ measurements on the Argenti\`ere glacier (Mont-Blanc massif, French Alps) are used to evaluate the EBM and the data assimilation scheme. Results indicate that snow stratigraphic profiles obtained after the analysis process show a closer agreement with the measured ones than the initial ones, and therefore demonstrate the high potential of assimilating SAR data to model of snow evolution.
\end{abstract}

\IEEEpeerreviewmaketitle
\section{Introduction}

\IEEEPARstart{S}{nowpack} characterization has become a critical issue in the present context of climate change. Estimating some of the properties of a snowpack, like its density and grain size distribution will provide great benefit to snow forecasting, prevision of natural hazard, like snow avalanche warning, and economic arrangements related to tourism and winter sports. Due to its imaging capabilities over large areas, unaffected by weather and day-night conditions, Synthetic Aperture Radar (SAR) is an important tool for snowpack characterization in a natural environment. Moreover, the high penetration depth of radar electromagnetic waves allow us to retrieve the information inside the volume of the snowpack. Over the past decade, the large availability of L and C-band SAR data provided by various spaceborne sensors, like ALOS PALSAR, ERS-1, ENVISAT, led to many studies on the characterization of snowpack properties~\cite{Shi-00a,Long-09}.

A new generation of X-band (8-12GHz) SAR systems, and in the near future Ku-band (12-18GHz), with high image resolution, short revisit time will provide improved information that might be used to characterize and monitor snowpack. In this context, it is necessary to develop a compatible EBM accounting for electromagnetic waves (EMW) propagation and scattering at high frequencies (X and Ku-bands) through a multilayer snowpack. Some backscattering models at L and C-band frequencies have been introduced in~\cite{Long-09,Kosk-10}. These models simulate the loss of EMW energy while propagating through dense media by solving the Radiative Transfer (RT) differential equation~\cite{Ulab-81c}. In order to introduce coherent recombination effects in the RT coherent model, Wang. et al.~\cite{Wang-00} applied the Strong Fluctuation Theory (SFT) introduced by Stogryn~\cite{Stog-84} to calculate the effective permittivity of each snow layer, in which the correlation among particles was taken into account. The scattering and absorption mechanisms in the EBM are simulated using the Rayleigh scattering model due to the snow grain size being in this study is much smaller than the carrier wavelength.

In this paper, the snowpack backscattering model initially developed in~\cite{Long-09} is adapted for X-band and higher frequencies, in the case of a dry snow medium. The adaptation consists of updating the IEM introduced by Fung et. al. in 1992~\cite{Fung-92} by a newer version published in 2004~\cite{Fung-04}, which allows the calculation of surface and ground backscattering components for X-band and higher frequencies. Meanwhile the modeling of volume backscattering of the existed model, which is based on solving the Vector Radiative Transfer equation and Rayleigh scattering model, is compatible for X and Ku-bands. From the physical features of each snow layer (optical grain radius, density, thickness) and for given SAR acquisition conditions (frequency, incidence angle), the model calculates the total backscattering coefficient $\sigma^0_{pq}$ for different polarization channels and their vertical distribution within the snowpack. Next, the snowpack profiles generated by the detailed snowpack model Crocus using downscaled meteorological fields from the SAFRAN analysis~\cite{Dura-93,Dura-09,Vion-12}, are constrained using the SAR image data and EBM simulations. In this study, the number of observable, i.e. the SAR backscattering coefficients, being much smaller than the number of unknown parameters, i.e. the snow cover properties, a classical estimation approach based on the use of an inverse problem would reveal totally inefficient. Instead, an adjoint operator of the direct EBM is developed to be used in a assimilation scheme. A variational assimilation method allows the integration of the observation data into a set of initial guess parameters through a direct model, and therefore can constrain these parameters without explicitly inverting the model. In our study, the three-dimensional variational analysis (3D-VAR) method~\cite{Cour-98} is implemented. Finally, a time series of TerraSAR-X acquisitions on the mountainous region of the French-Alps is used to evaluate the model and the data assimilation process. The Argenti\`ere glacier area has been chosen for the case study due to its large, uniformly snow-covered surface area. Some in-situ measurements on this area are also available at the same timeline of SAR acquisitions and therefore are used to evaluate the EBM and the performance of the data assimilation scheme.

Details of the EBM equations and its the physical and mathematical hypothesis are presented in section II. An introduction to the Crocus detailed snowpack model and to the detailed implementation of the 3D-VAR scheme are described in section III. Section IV shows a description of TerraSAR-X acquisition parameters, as well as results of the case study on the Argenti\`ere glacier.

\section{EMW backscattering model}

\subsection{Main components of the total backscattering coefficient}
\begin{figure}[!t]
   \centering
 \includegraphics[scale=0.45]{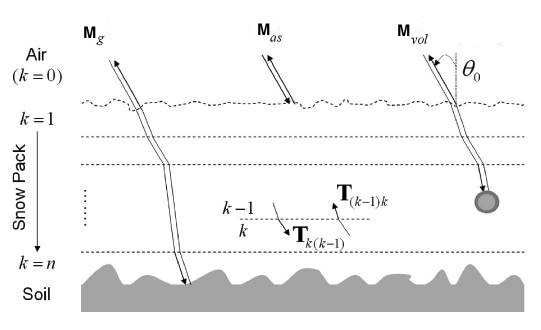}
\caption{Main backscattering mechanisms occurring within a multilayer snowpack that can be simulated using the RT theory at order 1: air-snow reflection ($\textbf{M}_{as}$), volume scattering ($\textbf{M}_{vol}$) and reflection over the ground ($\textbf{M}_g$).}\medskip
\label{fig:snow}
\end{figure}

The Stoke vector, which contains the incoherent information related to the polarization of an electromagnetic wave, can be expressed  as:
\begin{equation}
\textbf{g} =    \begin{bmatrix} 
                                                	\left<\left|E_h\right|^2\right> +  \left<\left|E_v\right|^2\right> \\
						\left<\left|E_h\right|^2\right> -  \left<\left|E_v\right|^2\right> \\
						 2\Re e\left<E_hE_v^*\right> \\
						 -2\Im m\left<E_hE_v^*\right> 
                                                    \end{bmatrix}
\end{equation}
where $E_h$ and $E_v$ represent the horizontal and vertical components of the Jones vector on the electric field~\cite{Lee-09}, and $\left<.\right>$ represents the expectation operator.

For given acquisition conditions, the Stoke vector scattered by a medium, $\textbf{g}_s$, can be related to the incident one, $\textbf{g}_i$, by a Mueller matrix \textbf{M} as $\textbf{g}_s = \textbf{Mg}_i$ with:
\begin{equation}
\textbf{M} = 
   \begin{bmatrix} 
	M_{11} & 0 & 0 & 0 \\
	0 & M_{22} & 0 & 0 \\
	0 & 0 & M_{33} & M_{34} \\
	0 & 0 & -M_{34} & M_{33}
   \end{bmatrix}
\label{eq:mueller}
\end{equation}
where $M_{11} = \sigma^0_{vv}$ and $M_{22} = \sigma^0_{hh}$ represent the co-polarized backscattering coefficients and $M_{33} = \Re e(\sigma^0_{vvhh})$ and $M_{34} = \Im m(\sigma^0_{vvhh})$ are correlation terms. Due to the reflection symmetry, the cross-polarization coefficients of the matrix $\sigma_{pppq}$ are equal to zero~\cite{Lee-09}, and the rest of the elements of \textbf{M} are null. 

The solution of the RT equation at order 1 provides a total backscattered information from a snowpack that consists of a combination of five scattering mechanisms: reflection at the surface air-snow interface, volume scattering, volume-ground and ground-volume interactions, and reflection over the ground~\cite{Mart-03}. Due to their small amplitude, the volume-ground and ground-volume contributions can be neglected~\cite{Flor-01}. The illustration of the three other mechanisms is shown in Fig.~\ref{fig:snow}. The expression of the total polarimetric backscattered information can be written using the Mueller matrix corresponding to each mechanism:
\begin{equation}
\textbf{M}_{snow} = \textbf{M}_{as} + \textbf{M}_{vol} + \textbf{M}_{g}
\label{eq:m_snow}
\end{equation}

The surface and ground backscattering are modeled using the IEM introduced by Fung et. al.~\cite{Fung-04}, whereas the volume contribution is calculated using the Vector Radiative Transfer equation.

\subsection{Surface backscattering}

The matrix $\textbf{M}_{as}$ represents the second order polarimetric response backscattered by the air-snow interface. Its elements can be calculated from the surface roughness parameters, e.g. its correlation function $w(x)$ and its root mean square (rms) height $\sigma_h$, the incidence angle $\theta_0$ and the emitted EM wave frequency $f$ using the Integration Equation Model (IEM)~\cite{Fung-04}. According to the IEM, the reflectivity may be expressed as: 
\begin{equation}
 \sigma^0_{pq} = \frac{k_0^2}{4\pi}\exp(-2k_0^2\sigma_h^2\cos^2\theta_0)\sum_{n=1}^{\infty}|I^n_{pq}|^2\frac{W^n(2k_0\sin\theta_0,0)}{n!}
\label{eq:sig_as}
\end{equation}
where $p$ and $q$ are equal to $h$ or $v$, indicating a horizontal or vertical polarization, $k_0=\frac{2\pi f}{c}$ represents the wave number. The detailed mathematical expressions of the surface spectrum $W^n$ and the Fresnel reflection/transmission factor $|I^n_{pq}|$ can be found in~\cite{Fung-04}.

\subsection{Volume backscattering}

The volume backscattering $\textbf{M}_{vol}$ is deduced from the loss of EMW intensity during propagation through a multilayer snowpack, which can be categorized into 4 types: related to transmission between two layers, absorption by the snow particles, scattering and coherent recombination. The amplitude of each mechanism depends largely on the dielectric properties of the snowpack medium. Therefore the permittivity of each layer, which characterizes its dielectric properties, needs to be calculated first:

\subsubsection{Dry snow permittivity}
Dry snow is considered as a dense and heterogeneous medium with strong variations of various physical properties such as grain size, density, thickness. Therefore the variance of permittivity across a snow layer is relatively high. Several snowpack characterization methods~\cite{Hall-86} are largely based on the assumption that the scattering losses due to the correlation of EMW are negligible. However, at high frequency, the snowpack structure becomes bigger compared to the wavelength of X or Ku-band EM waves~\cite{Huin-99}. The correlation between particles can no longer be ignored. The Strong Fluctuation Theory (SFT) introduced by Stogryn~\cite{Stog-84} can model the permittivity of such medium by using the effective permittivity ($\epsilon_{eff}$) that takes into account the scattering effect among ice particles at high frequencies. The expression of $\epsilon_{eff}$ using SFT is as follows~\cite{Huin-99}:
\begin{equation}
 \epsilon_{eff} = \epsilon_g + j.\frac{4}{3}\delta_{\epsilon_g}.k_0^3.\sqrt{\epsilon_g}.L^3
\end{equation}
where $\epsilon_g$ and $\delta_{\epsilon_g}$ are the quasi-static permittivity and its variance, $k_0$ the wave number and $L$ the correlation length, which is proportional to the average snow grain size and the snow density of the medium.

\subsubsection{Transmission between two layers}

The snowpack consists of layers with different physical properties. Therefore the model needs to take into account the energy loss due to transmission between two layers. With the assumption of a smooth interface between two layers, the Fresnel transmission can be used. It is expressed through a matrix as follows~\cite{Ulab-81c}:

\begin{equation}
 \textbf{T}_{k(k-1)} = \frac{\epsilon_{k-1}}{\epsilon_k}\begin{bmatrix}
					\left|t^{vv}_{k(k-1)}\right|^2 & 0 & 0 & 0 \\
					0 & \left|t^{hh}_{k(k-1)}\right|^2 & 0 & 0 \\
					0 & 0 & g_{k(k-1)} & -h_{k(k-1)} \\
					0 & 0 & h_{k(k-1)} & g_{k(k-1)}
				\end{bmatrix}
\label{eq:t}
\end{equation}
where $\left|t^{pp}_{k(k-1)}\right|^2$ represents the Fresnel transmission coefficients of $pp$ channel, whereas $g_{k(k-1)}$ and $h_{k(k-1)}$ are the terms of Mueller matrix related to the co-polarized correlation~\cite{Long-09}: 
\begin{equation}
g_{k(k-1)} = \frac{\cos\theta_{k-1}}{\cos\theta_k}\Re e(t^{vv}_{k(k-1)}t^{hh*}_{k(k-1)})~\text{and}~
h_{k(k-1)} = \frac{\cos\theta_{k-1}}{\cos\theta_k}\Im m(t^{vv}_{k(k-1)}t^{hh*}_{k(k-1)}) 
\end{equation}

\subsubsection{The attenuation}

The particles in a snowpack are generally considered as spheres~\cite{Flor-01,Long-08,Kosk-10}. Due to the spherical symmetry of the particle shape, the extinction of a wave propagating through the snowpack is independent of the polarization and may hence be represented by a scalar coefficient. The extinction is composed of an absorption and a scattering terms:
\begin{equation}
 \kappa_e = \kappa_a + \kappa_s
\end{equation}
It can also be computed through the effective permitivity $\epsilon_{eff}$~\cite{Huin-99}:
\begin{equation}
 \kappa_e = 2k_0Im\left|\sqrt{\epsilon_{eff}}\right|
\end{equation}

The attenuation matrix represents the gradual loss in EMW intensity while penetrating through a multilayer snowpack, composed of layers with different physical properties. It takes into account the energy loss by absorption and scattering mechanisms based on the extinction coefficient $\kappa_e$ and thickness $d$ of the layer, as well as the loss by transmission effect while an EM propagate through different layers:
\begin{eqnarray} 
\textbf{Att}_{down}(k) = \prod_{i=1}^k\text{exp}\left(-\frac{\kappa_e^id^i}{\cos\theta_i}\right)~\textbf{T}_{i(i-1)} \\
\textbf{Att}_{up}(k) = \prod_{i=1}^k\textbf{T}_{(i-1)i}~\text{exp}\left(-\frac{\kappa_e^id^i}{\cos\theta_i}\right) 
\end{eqnarray}
The $\textbf{Att}_{down}$ is the intensity loss when propagating from the surface to layer $k$, whereas $\textbf{Att}_{up}$ represents the intensity loss from layer $k$ to the surface. The exponential factor, which takes into account the gradual loss of energy throughout the layer, is deduced from the basic radiative transfer equation $dI = I\kappa_e dr$ where $r = d/cos\theta$. 

\subsubsection{Scattering by the particles}

The phase matrix $\textbf{P}^k$ under the hypothesis of spherical particles has the form shown in (\ref{eq:mueller}) where the cross-polarization terms $P_{12}$ and $P_{21}$ are null. In the backscattering case, with the assumption of spherical particles, the SFT phase matrix can be simplified to $\textbf{P}^k = \frac{3\kappa_s}{8\pi}I_4$ where $I_4$ is the (4x4) identity matrix~\cite{Tsan-07}.

\subsubsection{Calculation of the volume backscattering}

If we consider a snowpack made of $n$ distinct layers, where $\theta_k$ is the incidence angle and $d^k$ is the thickness of layer $k$, the total contribution of the volume backscattering mechanism $\textbf{M}_{vol}$ can be written as follows:

\begin{align}
\textbf{M}_{vol} = &~4\pi\cos\theta_0\sum_{k=1}^{n}\textbf{Att}_{up}(k-1)\textbf{T}_{(k-1)k} \nonumber \\ &.\frac{1-\text{exp}\left(-\frac{2\kappa_e^kd^k}{\cos\theta_k}\right)}{2\kappa_e^k}\textbf{P}^k\textbf{T}_{k(k-1)}\textbf{Att}_{down}(k-1)
\label{eq:sig_vol}
\end{align}

\subsection{Ground backscattering}

The backscattering $\textbf{M}_g$ of the snow-ground interface may be computed as:
\begin{equation}
\textbf{M}_g = \cos\theta_0~\textbf{Att}_{up}(n)\frac{\textbf{R}(\theta_n)}{\cos\theta_n}\textbf{Att}_{down}(n)
\label{eq:sig_g}
\end{equation}
where $\textbf{R}(\theta_n)$ represents the contribution of the underlying ground surface backscattering and can be determined using the IEM.

\section{3d-var data assimilation}

\subsection{The detailed snowpack model Crocus}
\begin{figure}[!t]
   \centering
\includegraphics[scale = 1]{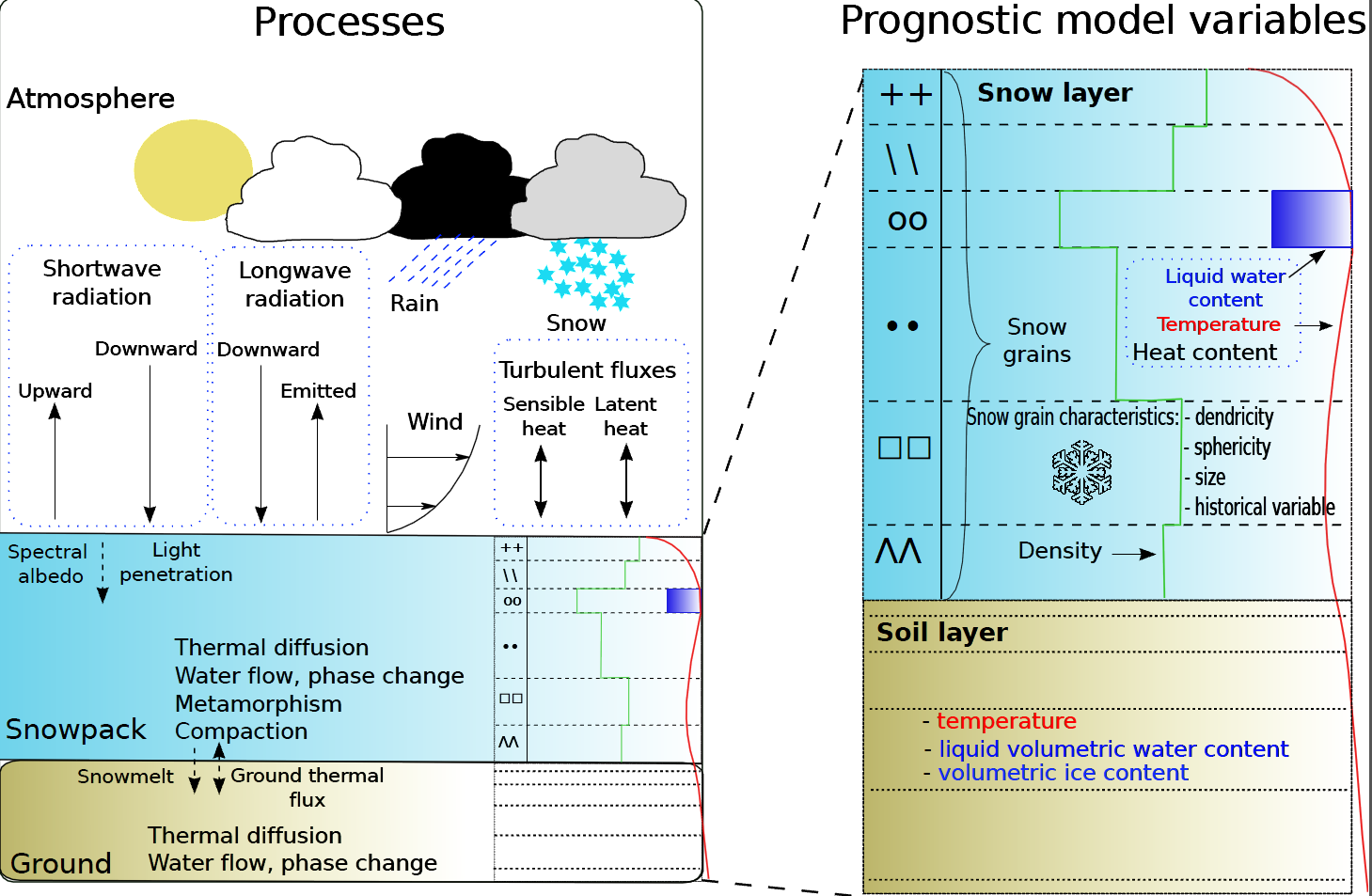}
\caption{General scheme of Crocus processes and variables .}\medskip
\label{fig:crocus}
\end{figure}

Crocus is a one-dimensional numerical model simulating the thermodynamic balance of energy and mass of the snowpack. Its main objective is to describe in detail the evolution of internal snowpack properties based on the description of the evolution of morphological properties of snow grains during their metamorphism. Fig.~\ref{fig:crocus} describes the general scheme of Crocus. It takes as input the meteorological variables air temperature, relative air humidity, wind speed, solar radiation, long wave radiation, amount and phase of precipitation. When it is used in the French mountain ranges (Alps, Pyrenees and Corsica), these quantities are commonly provided by the SAFRAN system, which combines ground-based and radiosondes observations with an \emph{a priori} estimate of meteorological conditions from a numerical weather prediction (NWP) model~\cite{Dura-93,Dura-09}. The output includes the scalar physical properties of the snowpack (snow depth, snow water equivalent, surface temperature, albedo, \dots) along with the internal physical properties for each layer (density, thickness, optical grain radius, \dots). SAFRAN meteorological fields are assumed to be homogeneous within a given mountrain range and provide a description of the altitude dependency of meteorological variables by steps of 300 m elevation~\cite{Dura-93, Dura-09}.

Here we use the latest version of the detailed snowpack model Crocus, recently incorporated in the land surface scheme ISBA within the SURFEX interface~\cite{Vion-12}. Among other advantages over previous versions of Crocus, this allows seamless coupling of the snowpack to the state of the underlying ground.
\subsection{Method introduction}

\begin{figure}[!t]
   \centering
\includegraphics[width = 3.2in]{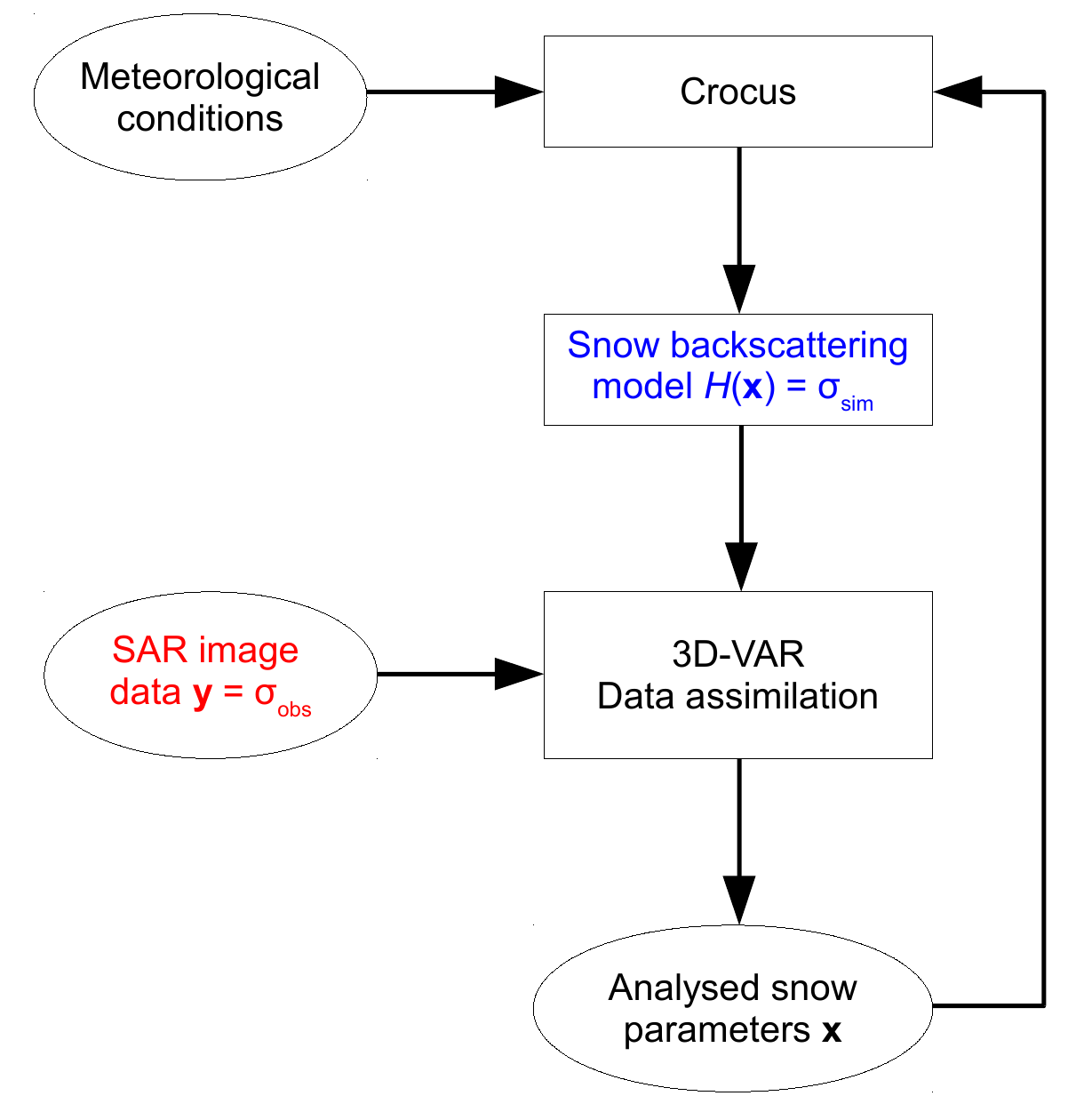}
\caption{Global scheme of the data assimilation used in this study. The input of the process are the SAR reflectivities, $\sigma^0$, (observation) and the snowpack stratigraphic profile calculated by Crocus (guess). The output is the assimilated snowpack profile $\textbf{x}$ that minimizes the cost function.}\medskip
\label{fig:schema}
\end{figure}
Variational assimilation aims to integrate observation data into guess parameters through the use of an observation operator. It is widely used in meteorological studies in order to relate observations, measurements and modeling aspects~\cite{Dumo-12}. The method concentrates on searching a solution that minimizes simultaneously the distance between observations and simulation results and the distance between initial guess variables and the analysed variables. A scheme of this process is presented in Fig.~\ref{fig:schema}. In this part of the paper, the output of our EBM in the previous section, such as backscattering coefficient of HH and VV polarizations, are used as elements of the observation operator $\textbf{H}_{ebm}(\textbf{x})$:
\begin{equation}
\textbf{H}_{ebm}(\textbf{x}) = \text{vec}(\textbf{M}_{snow})
\end{equation}
where \textbf{x} represents the set of variables required to describe the snowpack properties.

The 3D-VAR~\cite{Cour-98} algorithm is based on the minimization of a cost function $J(\textbf{x})$, defined as:
\begin{equation}
J(\textbf{x})=(\textbf{x}-\textbf{x}_g)^t\textbf{B}^{-1}(\textbf{x}-\textbf{x}_g)+(\textbf{y}_{obs}-\textbf{H}_{ebm}(\textbf{x}))^t\textbf{R}^{-1}(\textbf{y}-\textbf{H}_{ebm}(\textbf{x}))
\end{equation}
where $\textbf{x}$ is called the state vector, and can be modified after each iteration of minimization, $\textbf{x}_g$ is the initial guess of the state vector and remains constant during the whole process. Therefore $\|\textbf{x} - \textbf{x}_g\|^2$ serves as a distance between the modified profile and the starting point. The observed polarimetric response, $\textbf{y}_{obs}$, is denoted similarly to the calibrated values of the backscattering coefficients $\sigma^0$. Therefore, $\|\textbf{y} - \textbf{H}_{ebm}(\textbf{x})\|^2$ represents the distance between simulated and observed quantities in the observation space. The process also requires the estimation of the error covariance matrices of observations/simulations $\textbf{R}$ and of the model $\textbf{B}$, the guess error variance.

\subsection{Adjoint operator and minimization algorithm}

In order to minimize the cost function $J$, one needs to calculate its gradient:
\begin{equation}
 \nabla J(\textbf{x})=\frac{\partial J(\textbf{x})}{\partial \textbf{x}} = 2\textbf{B}^{-1}(\textbf{x}-\textbf{x}_g)-2\nabla \textbf{H}_{ebm}^t(\textbf{x})\textbf{R}^{-1}(\textbf{y}_{obs}-\textbf{H}_{ebm}(\textbf{x}))
\end{equation}

If the model is denoted $\textbf{H}_{ebm}:\mathcal{B}\rightarrow\mathcal{R}$, with $\mathcal{B}$ and $\mathcal{R}$ are the domain of definition of \textbf{x} and \textbf{y}, then the function $\nabla \textbf{H}_{ebm}^t$ satisfying: $\forall \textbf{x},\textbf{y},~\langle \nabla \textbf{H}_{ebm}^t\textbf{y},\textbf{x}\rangle_\mathcal{B} = \langle \textbf{y},\nabla \textbf{H}_{ebm}\textbf{x}\rangle_\mathcal{R}$ is the adjoint operator of $\textbf{H}_{ebm}$.

Once the adjoint operator is developed, the minimization of J can be achieved using a gradient descent algorithm. Each iteration consists of modifying the vector $\textbf{x}$ by a factor according to the Newton method:
\begin{equation}
 \textbf{x}_{n+1} = \textbf{x}_n + (\nabla^2 J(\textbf{x}_n))^{-1}\nabla J(\textbf{x}_n)
\end{equation}
where $\nabla^2 J(\textbf{x}_n)$ is the gradient of second order (hessian) of $J$:
\begin{equation}
\nabla^2 J = 2\textbf{B}^{-1} + 2\nabla \textbf{H}_{ebm}^t\textbf{R}^{-1}\nabla \textbf{H}_{ebm}
\label{eq:step}
\end{equation}

\subsection{Discussion on the assimilation method}
In general, the aim of modeling the relation between the elements of natural environment and the observations measured by special equipments (such as SAR or optical sensors) is to try to inverse the model and estimate the variables of environment using the observations. However, such problems often lead to the need to resolve a underdetermined system, which means the number of unknown is higher than the number of equations. 

In our case, the length of the input state vector \textbf{x} can reach 100 (in the case of snowpack with 50 layers, which is frequently generated by Crocus), meanwhile the output of the model consists of only the backscattering coefficients corresponding to  polarimetric channels of SAR data. Therefore the realization of an inverse model is theoretically impossible. 

The data analysis method, on the other hand, requires a vector of guess variables relatively close to the actual values, allowing to add an \emph{a priori} information. The snowpack variables calculated by Crocus are used as guess in our assimilation scheme. The fundamental goal is to try to modify the initial guess variables, based on balancing the errors of guess, modeling and measurements. It should be noted that the problem stays underdetermined, the analysis scheme only serves as a method to improve the initial guess variables using the new observations from SAR data. The quality of improvement is based on the estimation of the initial guess vector $\textbf{x}_g$ and the precision of the EBM.
 
\section{Case Study: Argenti\`ere Glacier}

\subsection{Data}

\begin{table}[!t]
\caption{\emph{TerraSAR-X acquisitions parameters}}
\label{tab:data}
\vspace*{-0.5cm}
\renewcommand{\arraystretch}{1.5}
	\begin{center}
		\begin{tabular}{c|c}
		\hline
		\hline
		 \textbf{Parameter} & \textbf{Value} \\
		\hline
		TerraSAR-X products & Single Look Complex Image \\
		Frequency (GHz) & 9.65 \\
		Channels & HH\\
		Incidence angle (deg) & 37.9892\\
		Mode & Descending\\
		Acquisition dates & 6 Jan 2009, 17 Jan 2009, 28 Jan 2009 \\
		& 8 Feb 2009, 19 Feb 2009, 2 Mar 2009\\
		& 13 Mar 2009, 24 Mar 2009\\
		Range resolution (m) & 1.477\\
		Azimuth resolution (m) & 2.44\\
		Calibration gain (dB) & 49.6802\\ \hline
		\hline
		\end{tabular}
	\end{center}
\end{table}

\begin{figure}[!t]
   \centering
 \includegraphics[scale=0.28]{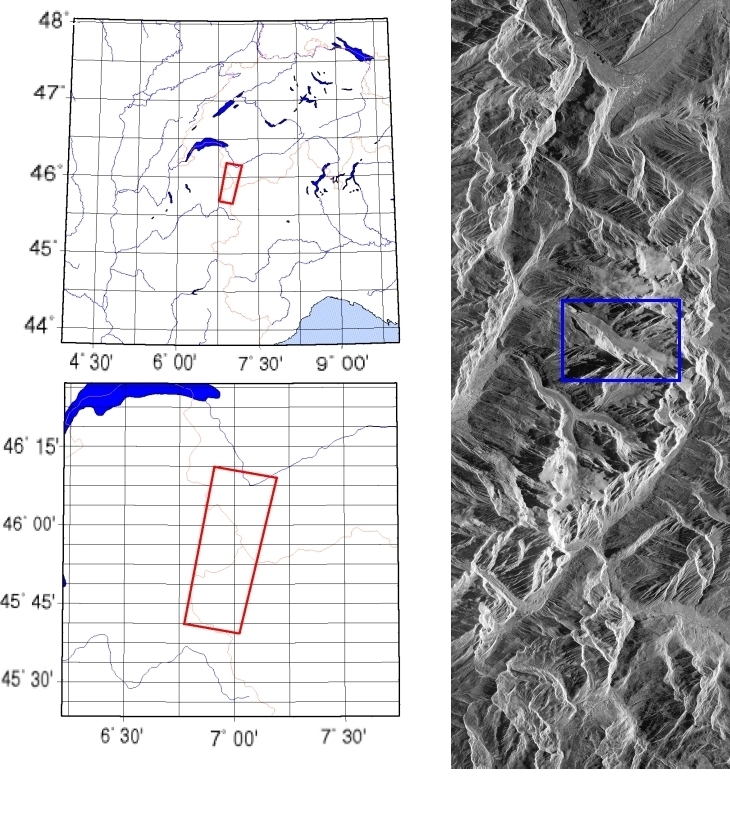}
 \includegraphics[width = 3.2in]{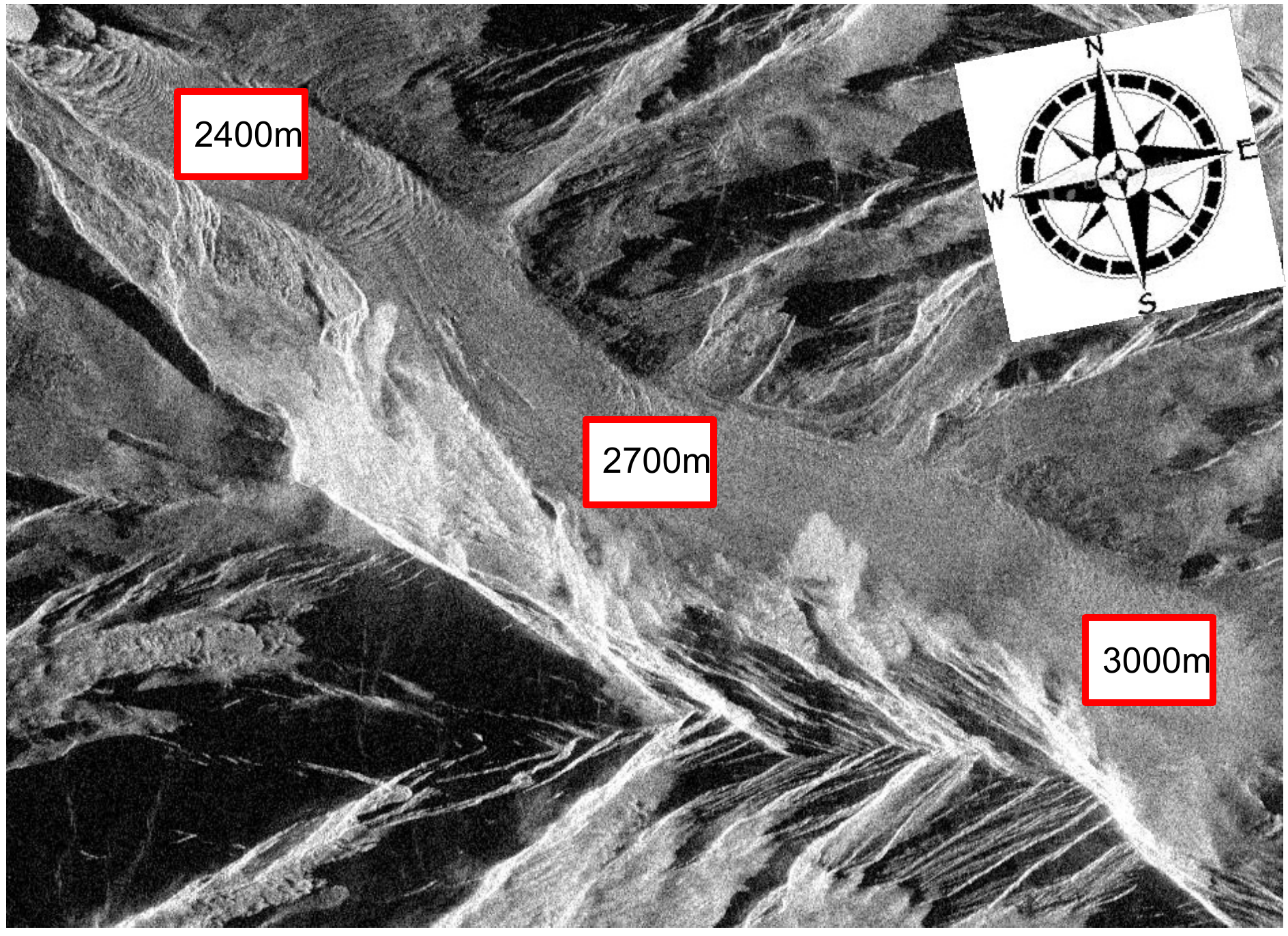}
\caption{(Top) Location of the TerraSAR-X acquisition in the French Alps. (Bottom) The crop image on the Argenti\`ere glacier area. The snowpack stratigraphic profiles calculated by Crocus snow model are given for 3 different altitudes on the Argenti\`ere glacier: 2400m, 2700m and 3000m. The rectangles show the approximate positions of these altitudes on the TerraSAR-X images.}
\label{fig:data}
\end{figure}

For this study, a time series of TerraSAR-X descending acquisitions on the region of Chamonix Mont-Blanc, France from 06 January 2009 to 24 March 2009 are available. A total of 8 SAR images are available every 11 days. Table~\ref{tab:data} shows the main parameters of TerraSAR-X data. The area of interest covers the Argenti\`ere glacier (Altitude: 2771m, 45.94628$^{\circ}$ N, 7.00456$^{\circ}$ E). The size of the domain is approximately 5km $\times$ 6km. Figure~\ref{fig:data} shows the location and the image of Argenti\`ere glacier captured on 06 January 2009. In order to obtain square pixels resolution, multi-look number of 5 for slant range and 3 for azimuth direction was applied. 

For this study, meteorological forcing data provided by SAFRAN at 2400, 2700, and 3000 m altitude on horizontal terrain were used to drive the detailed snowpack model Crocus throughout the  whole season 2008-2009 (starting on August 1st 2008). In order to carry out the comparison between the backscattering coefficients $\sigma_{sim}$ (obtained from executing the EBM using Crocus snowpack profile as input) and $\sigma_{TSX}$ (obtained from TerraSAR-X reflectivity), they need to be representative of the same area. Therefore we need to estimate the backscattering coefficients that well-represent the SAR reflectivities of the studied areas. The characteristic of a snowcover surface texture is spatially heterogeneous due to its strong variations of physical properties. A Gaussian distributed SAR texture hypothesis is therefore invalid. In recent studies, it has been proven that the texture of a SAR image of a heterogeneous medium can be modeled using the Fisher probability distribution~\cite{Bomb-08,Hara-11}. From the parameters of Fisher probability distribution, we can calculate the theoretical mean value which represents the backscattering coefficient of an area. In this study, the representative values of the backscattering coefficients of SAR image data on each altitude are obtained by calculating the mean value of Fisher-distributed texture of three regions (Fig. \ref{fig:data}) as in \cite{Hara-11}.

\begin{figure}[!t]
   \centering
\includegraphics[scale=0.5]{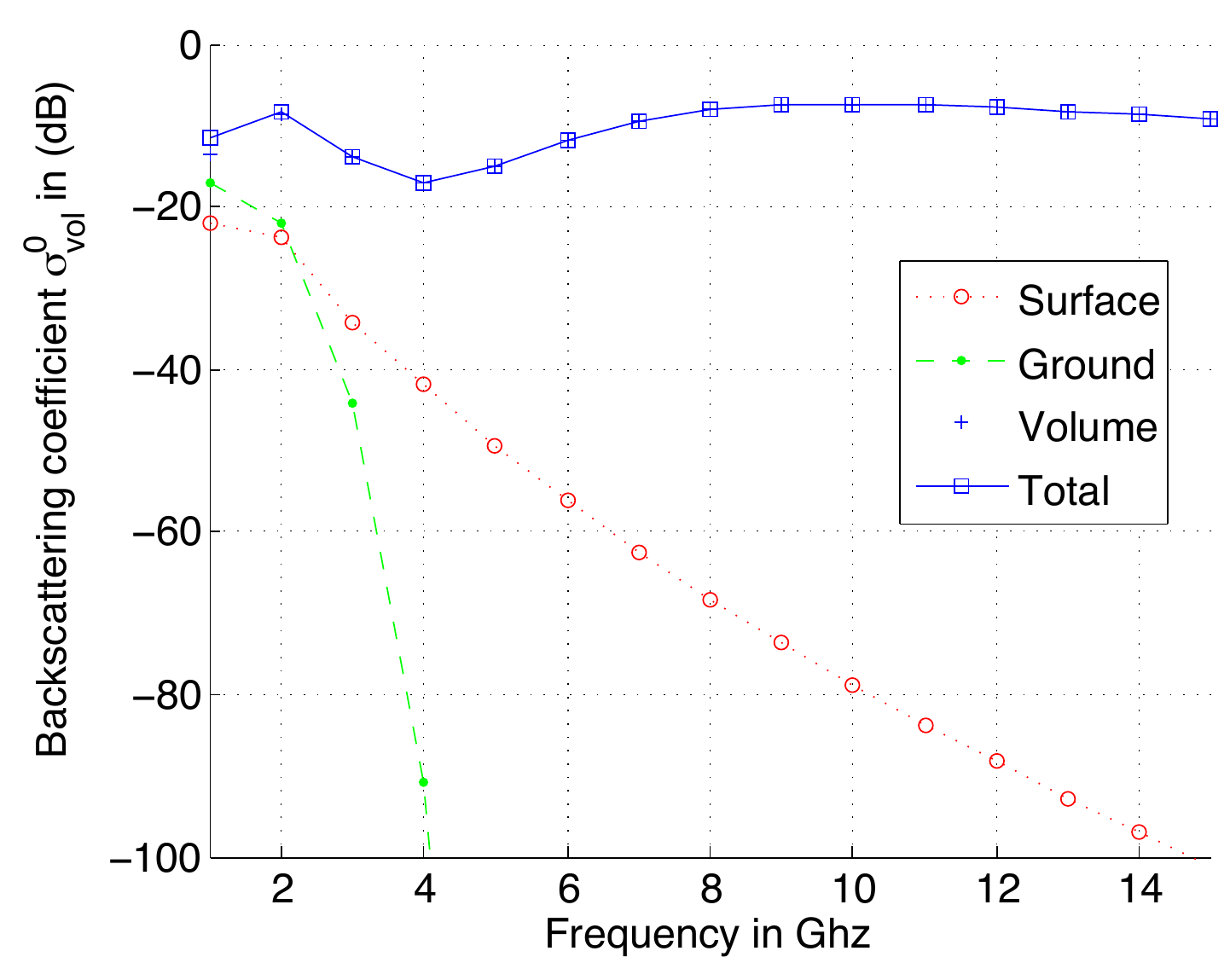}
\includegraphics[scale=0.5]{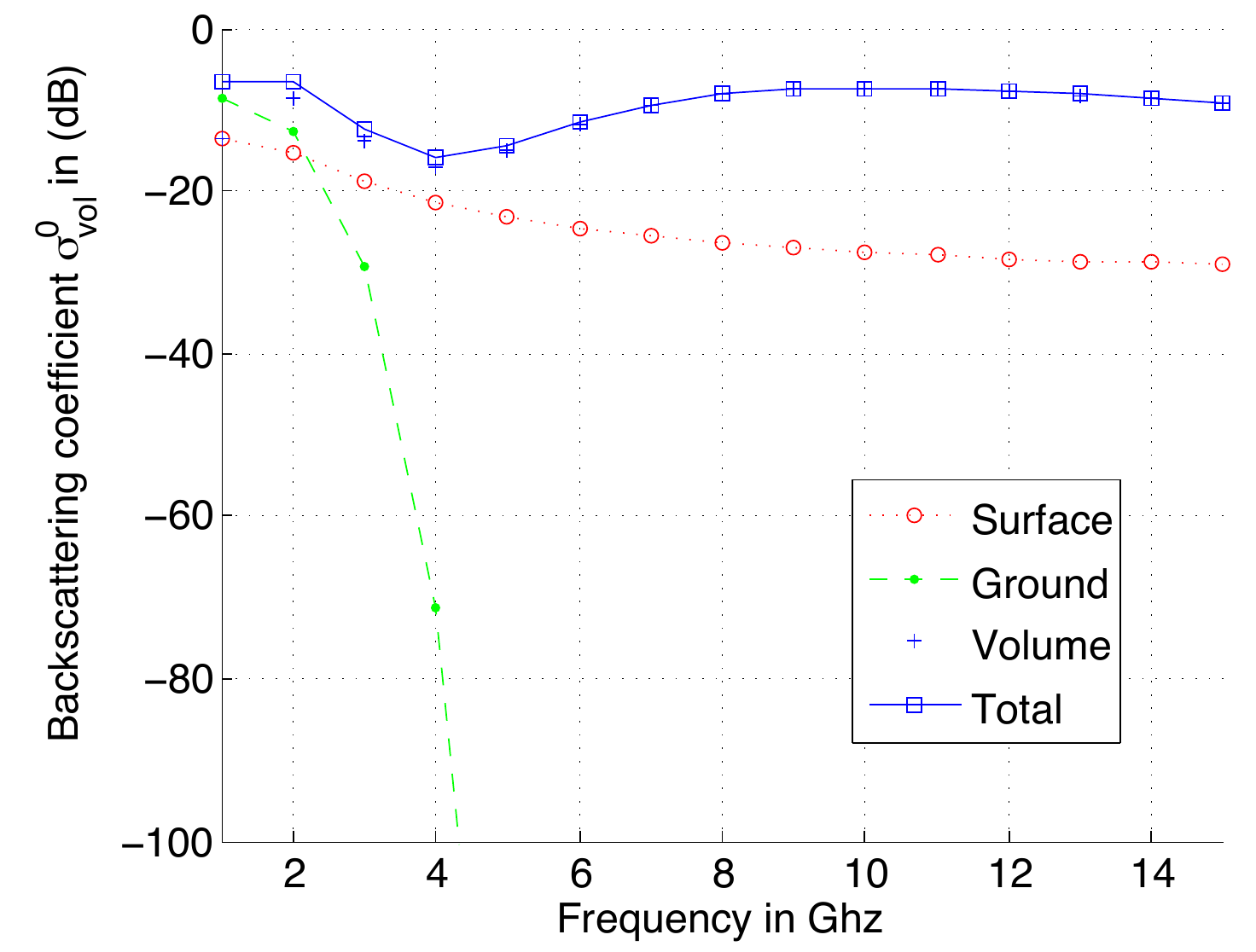}
\caption{Comparison of the contribution from surface, ground and volume backscattering mechanisms plotted in function of frequency $f$. The roughness parameters of surface and ground are: Left: $\sigma_h = 0.4cm$ and $l = 8.4cm$ corresponding to slightly rough and Right: $\sigma_h = 1.12cm$ and $l = 8.8cm$ corresponding to very rough (data taken from~\cite{oh-92}). The volume backscattering coefficient is simulated using the snow profile in Table~\ref{tab:pro_str}.}\medskip
\label{fig:sol_1}
\end{figure}

The roughness parameters of surface air-snow and ground are not available in the guess data calculated by Crocus, therefore the empirical values of the correlation length and the rms height have been taken from the measurements of Oh et. al.~\cite{oh-92}. As we can see in figure~\ref{fig:sol_1}, in the high frequency range (more than 10GHz), the contribution of surface and ground contribution are considerably low compared to the volume backscattering, regardless of slightly rough or very rough surface. Therefore we only concentrate on the volume contribution on the tests of data analysis method. According to the nature of a dry snow surface, the values of $\sigma_h = 0.4cm$ and $l = 8.4cm$ with a Gaussian type of surface spectrum, which correspond to a slightly rough surface, are used for the modeling in this study. With these surface and ground parameters set to constants, the original input vector $\textbf{x}= [\textbf{x}_{Crocus}~\textbf{x}_s~ \textbf{x}_g]$ in our case contains only the physical parameters of each layer of snowpack, which has the following form:
\begin{equation}
 \textbf{x} = [\textbf{x}_{Crocus}] = [x_1,x_2,\dots,x_{2n}]^t = [d_1,d_2,\dots,d_n,\rho_1, \rho_2,\dots,\rho_n]^t
\end{equation}
where $d_i$ and $\rho_i$ are respectively the optical grain size and the density of $i^{th}$ layer of the snowpack. On the first iteration of the algorithm, $\textbf{x} = \textbf{x}_g$ and $\textbf{x}_g$ is given by the Crocus snow profile.

The covariance matrix $\textbf{B}$, which represents the error of the input profile, i.e.~of the Crocus calculation, is a square $(2n\times2n)$ definite positive matrix. Each element of $\textbf{B}$ is computed as:
\begin{equation}
 \textbf{B}_{i,j} = \sigma_i.\sigma_j.\gamma_{ij}
\end{equation}
where $\sigma_i = \sqrt{E[(\varepsilon_i - \bar{\varepsilon_i})^2]}$ represents the standard deviation of error while calculating $x_i$. In our case, all element of \textbf{x} is estimated using the snow metamorphism model Crocus, therefore the variances of error are the same, which are experimentally estimated to 0.3 mm and 65 kg/m$^3$ for the optical grain size calculation error and density calculation error respectively. 

The coefficient $\gamma_{ij}$ represents the correlation between errors of $x_i$ and $x_j$ and are modelled as:
\begin{equation}
 \gamma_{ij} = \beta e^{-\alpha \Delta h_{ij}}
\label{eq:corr}
\end{equation}
 where $\Delta h_{ij}$ is the distance in cm between layer $i$ and layer $j$. The values of $\alpha$ and $\beta$ depend on different types of correlations and can be splitted into 3 cases:
\begin{itemize}
\item Correlation $d$ - $d$: $\alpha = 0.11$ and $\beta = 1$
\item Correlation $\rho$ - $\rho$: $\alpha = 0.13$ and $\beta = 1$
\item Correlation $d$ - $\rho$: $\alpha = 0.15$ and $\beta = 0.66$
\end{itemize}
These values are issued from an ensemble of slightly perturbated Crocus runs, i.e. obtained by differences in their meteorological inputs, over one winter season. The deviation between these runs, considered as elementary perturbations, have been then statistically studied and fitted with the eq.~\ref{eq:corr} model for the two considered variables and their crossed value.

In this case study, the SAR data is available for HH channel, therefore the error covariance matrix $\textbf{R}$ is a scalar which is equal to the variance of SAR image intensity on the studied area. The calculations of the variance on the three altitudes of Argenti\`ere glacier gives the average value of $\textbf{R} = 0.03$. Nevertheless, after testing with different values, it has been observed that the output of the analysis algorithm is not very sensitive to this error factor. The scalar multiplication of 10 to 20 times the values of $\textbf{R}$ doesn't show noticeable effect on the result.

\subsection{Results and Discussion}

\begin{table}[!t]
\caption{\emph{Snow stratigraphic profile obtained from an in-situ measurement on Argenti\`ere glacier on 30 January 2009 at the altitude of 2700m.}}
\label{tab:pro_str}
\renewcommand{\arraystretch}{1.25}
	\begin{center}
		\begin{tabular}{|c|c|c|c|}
		\hline
		\hline
		\textbf{Snow depth} & \textbf{Thickness} & \textbf{Grain size} & \textbf{Density} \\
		\textbf{(cm)} & \textbf{(cm)} & \textbf{(mm/10)} & \textbf{(Kg/m$^3$)} \\ \hline
		0 & & & \\ \hline
		13 & 13 & 5 & 210 \\ \hline
		25 & 12 & 5 & 290  \\ \hline
		31 & 6 & 5 & 310  \\ \hline
		56 & 25 & 7.5 & 220 \\ \hline
		85 & 29 & 7.5 & 300 \\ \hline
		92 & 7 & 15 & 340 \\ \hline
		125 & 33 & 15 & 430 \\ \hline
		135 & 10 & 15 & 370 \\ \hline
		190 & 55 & 15 & 430 \\
		\hline
		\hline
		\end{tabular}
	\end{center}
\end{table}

Crocus snow stratigraphic profiles have been computed for 3 different altitudes over the Argenti\`ere glacier, on the dates of the TerraSAR-X acquisitions. Two in-situ snowpack profiles measurements are also available at the altitude of 2700m on 30 January 2009 and 17 March 2009, and an example is shown in table~\ref{tab:pro_str}. The level of liquid water content per volume at the time and location of measurement is below 2 percent, which means the snowpack can be considered as dry snow. Fig.~\ref{fig:data} shows the approximate locations of each study area on the glacier. 

\begin{figure}[!t]
   \centering
 \includegraphics[scale=0.4]{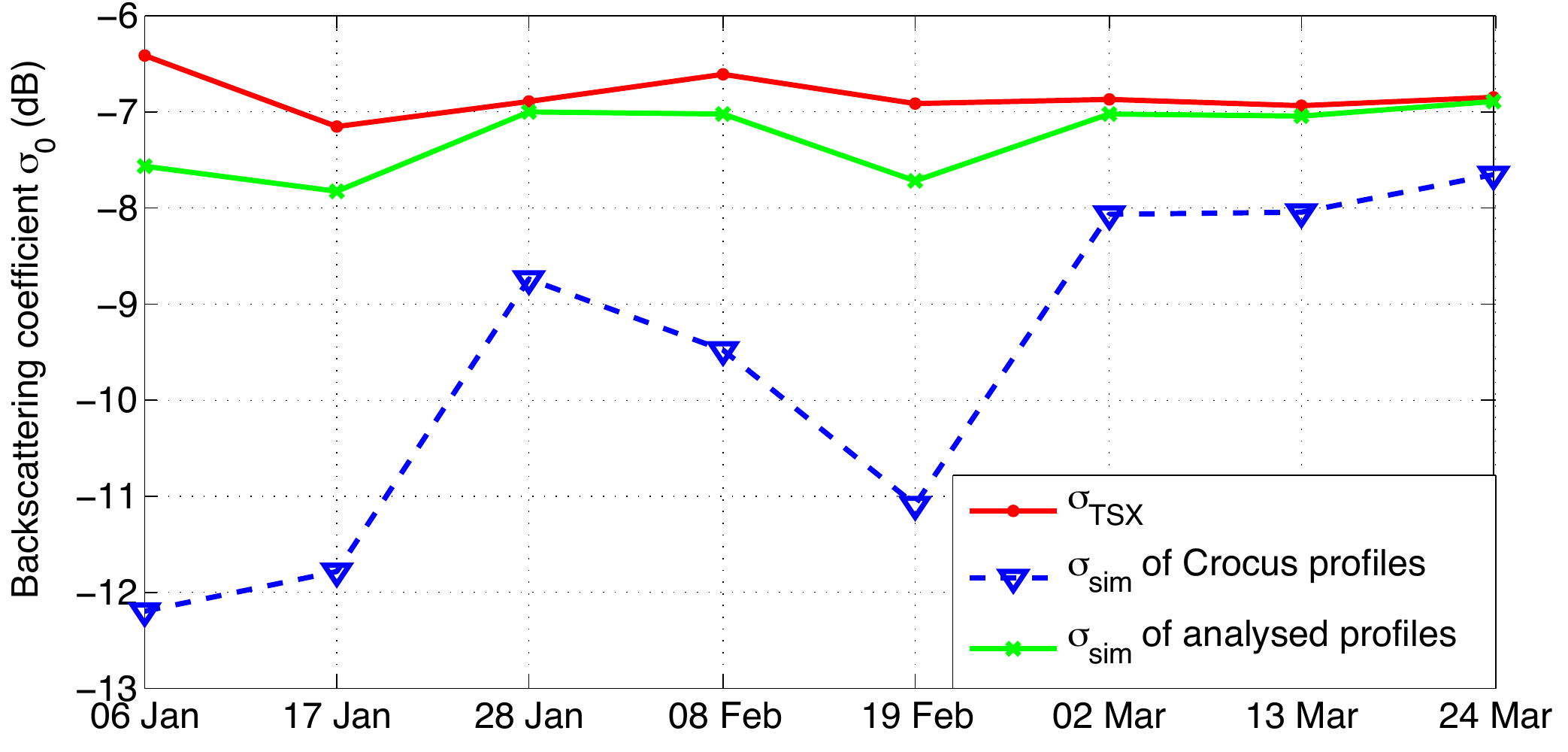}
\centerline{Altitude: 2400m}
\includegraphics[scale=0.4]{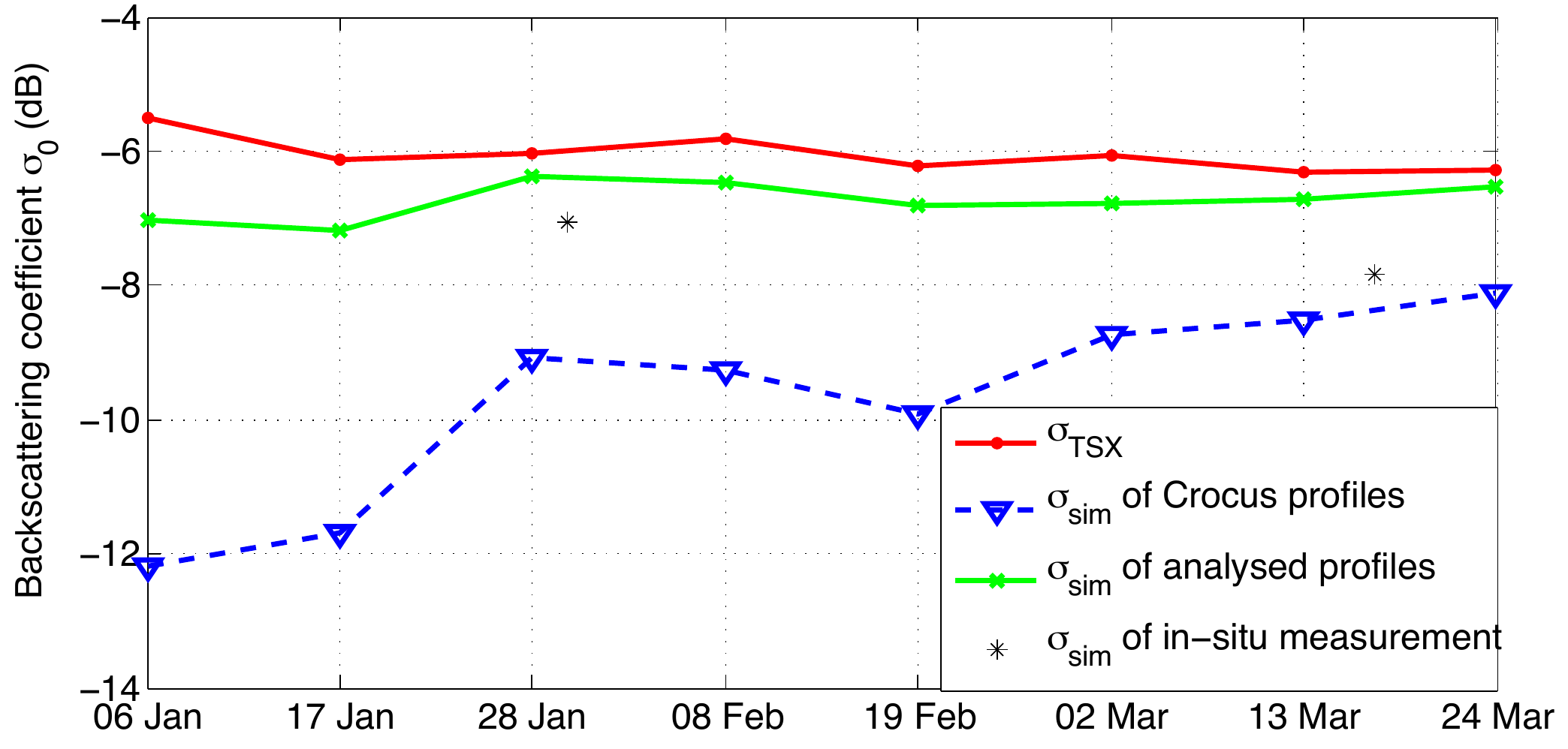}
\centerline{Altitude: 2700m}
\includegraphics[scale=0.4]{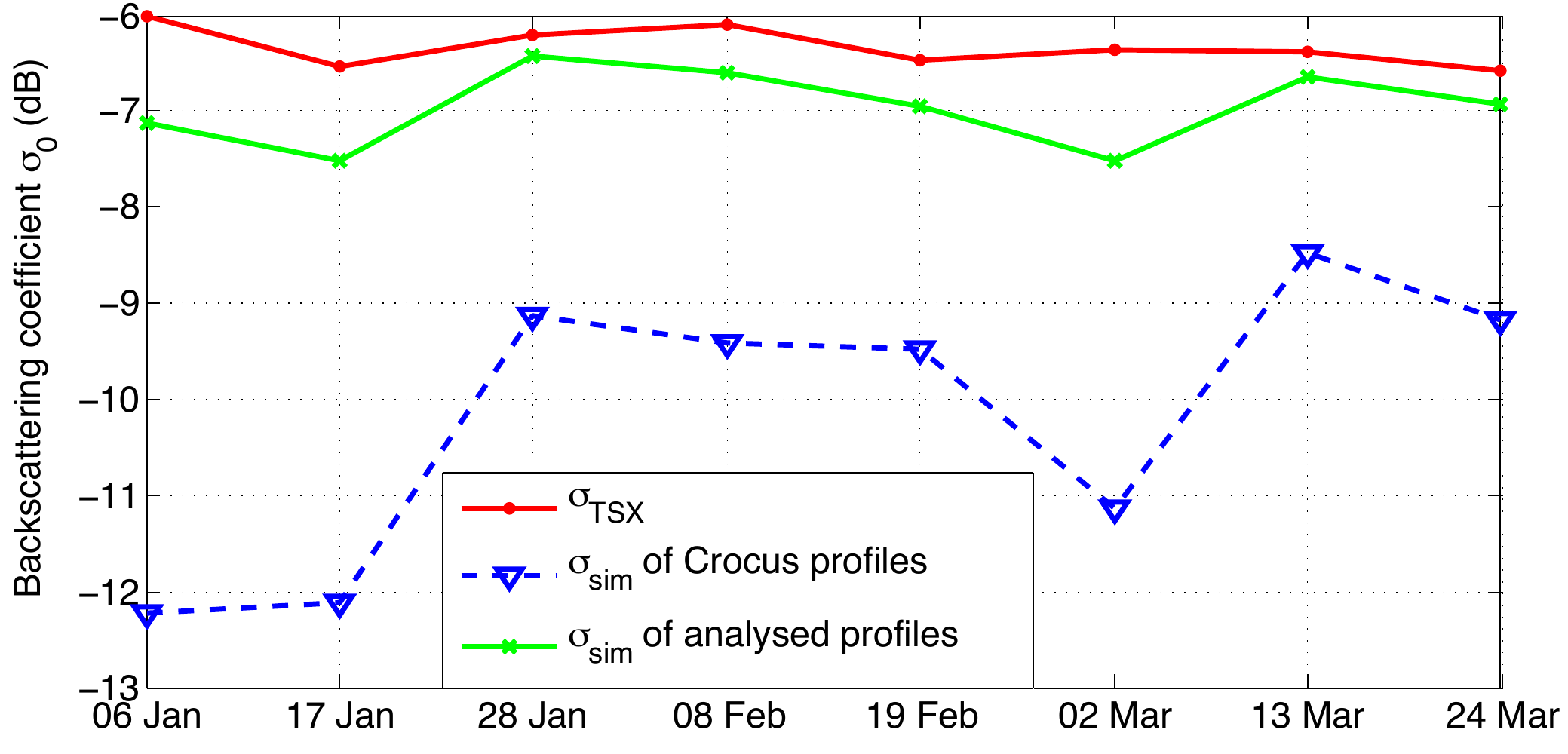}
\centerline{Altitude: 3000m}
\caption{Results of simulation and analysis using a time series of TerraSAR-X acquisitions in 2009 and the corresponding Crocus output . The $\sigma_{TSX}$ (red) are the mean values obtained from Fisher probability distribution on a region of SAR image. The $\sigma_{sim}$ (blue) represents the output of simulations using Crocus snowpack variables as input. The simulations again with the parameters after the data analysis process are shown in green.}\medskip
\label{fig:res1}
\end{figure}

\begin{figure*}[!t]
% \centerline{\subfloat{\includegraphics[width=2.3in]{2700_0106_d}%
% \label{fig_0106d}}
% \subfloat{\includegraphics[width=2.3in]{2700_0106_rho}%
% \label{fig_0106r}}
% \subfloat{\includegraphics[width=2.3in]{2700_0106}%
% \label{fig_0106}}}
% \centerline{Date: 6 Jan 2009. Altitude: 2700m.}
\centerline{\subfloat{\includegraphics[width=2.3in]{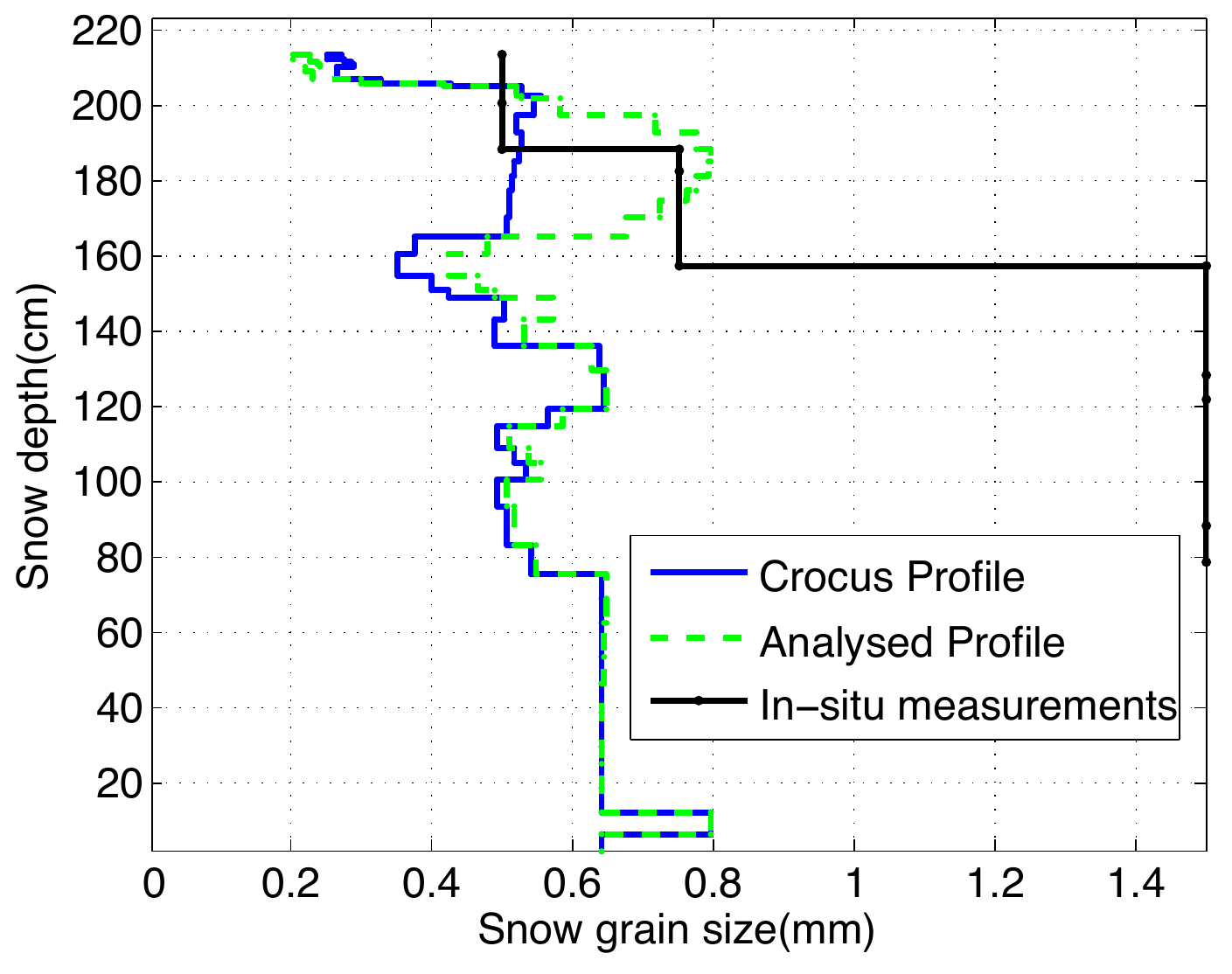}%
\label{fig_0128d}}
\subfloat{\includegraphics[width=2.3in]{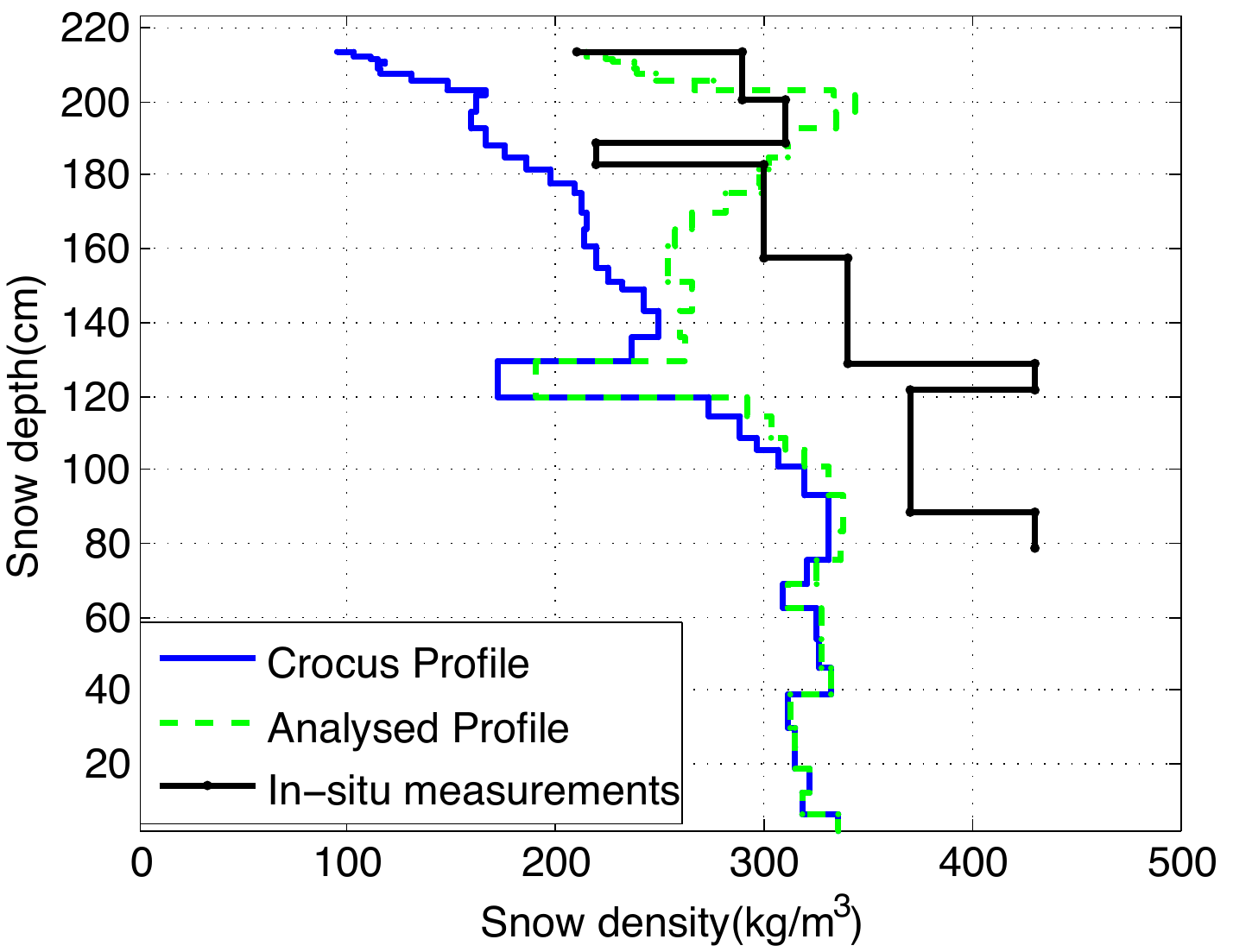}%
\label{fig_0128r}}
\subfloat{\includegraphics[width=2.3in]{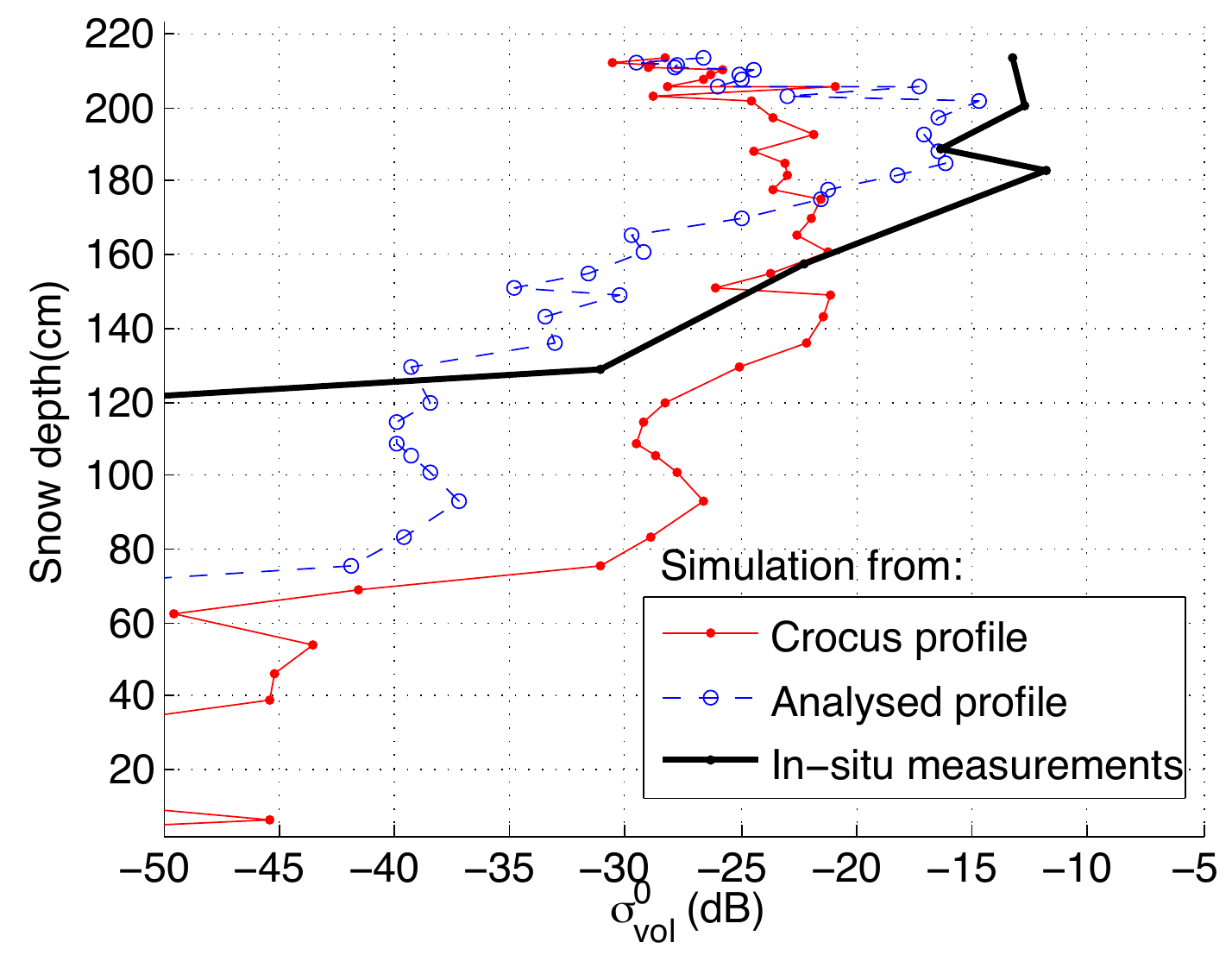}%
\label{fig_0128}}}
\centerline{Date: 28 Jan 2009. Altitude: 2700m.}
\centerline{\subfloat{\includegraphics[width=2.3in]{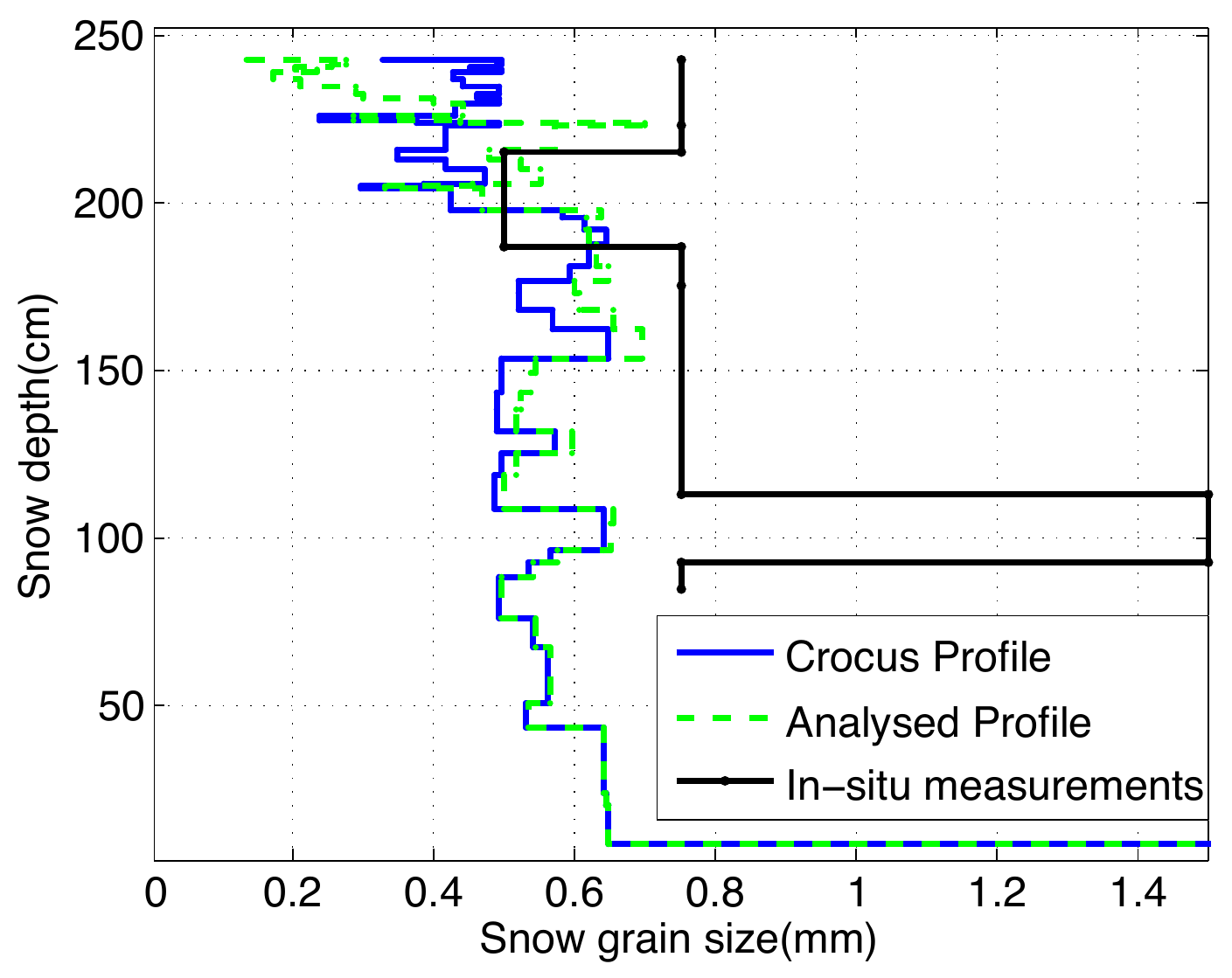}%
\label{fig_0313d}}
\subfloat{\includegraphics[width=2.3in]{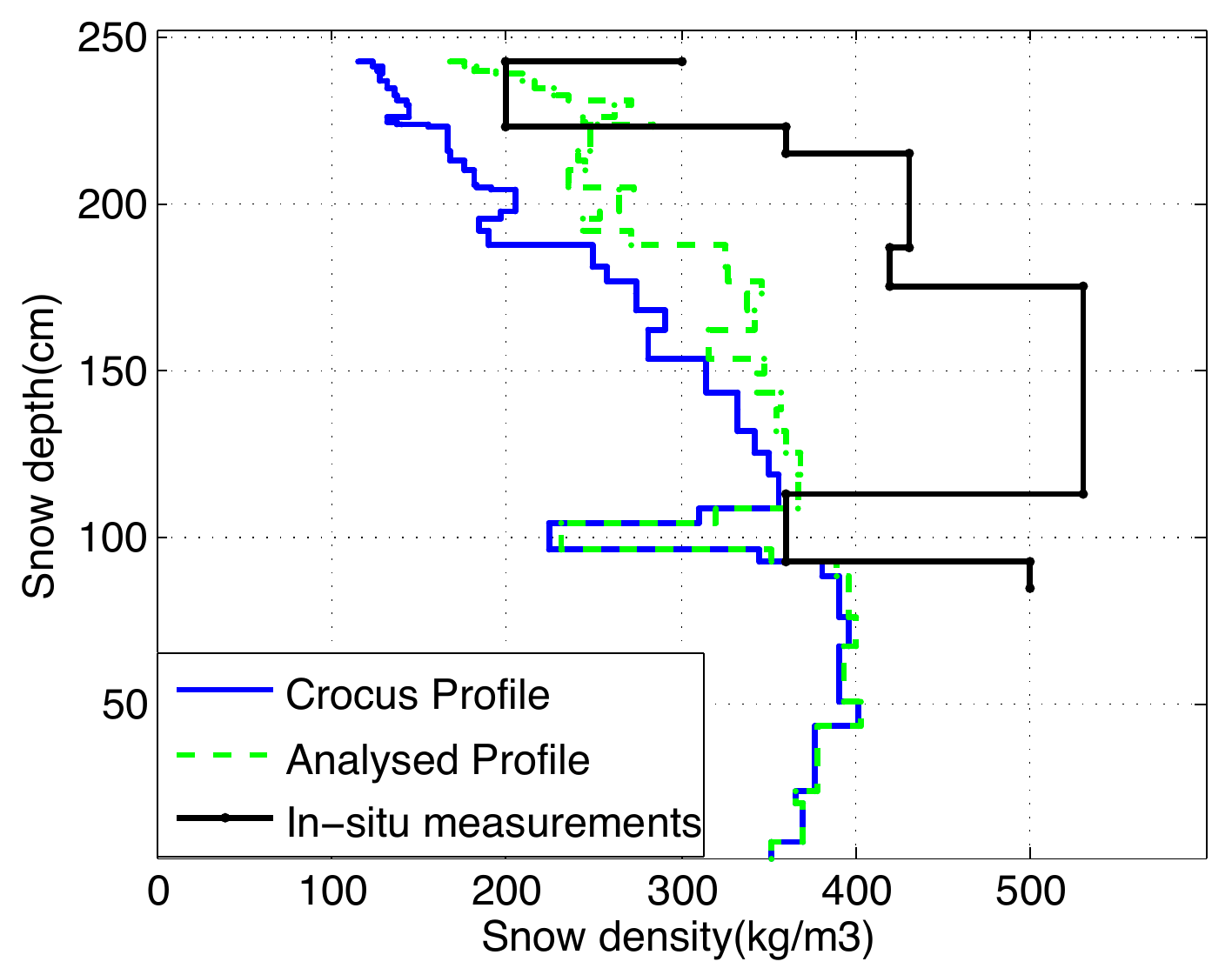}%
\label{fig_0313r}}
\subfloat{\includegraphics[width=2.3in]{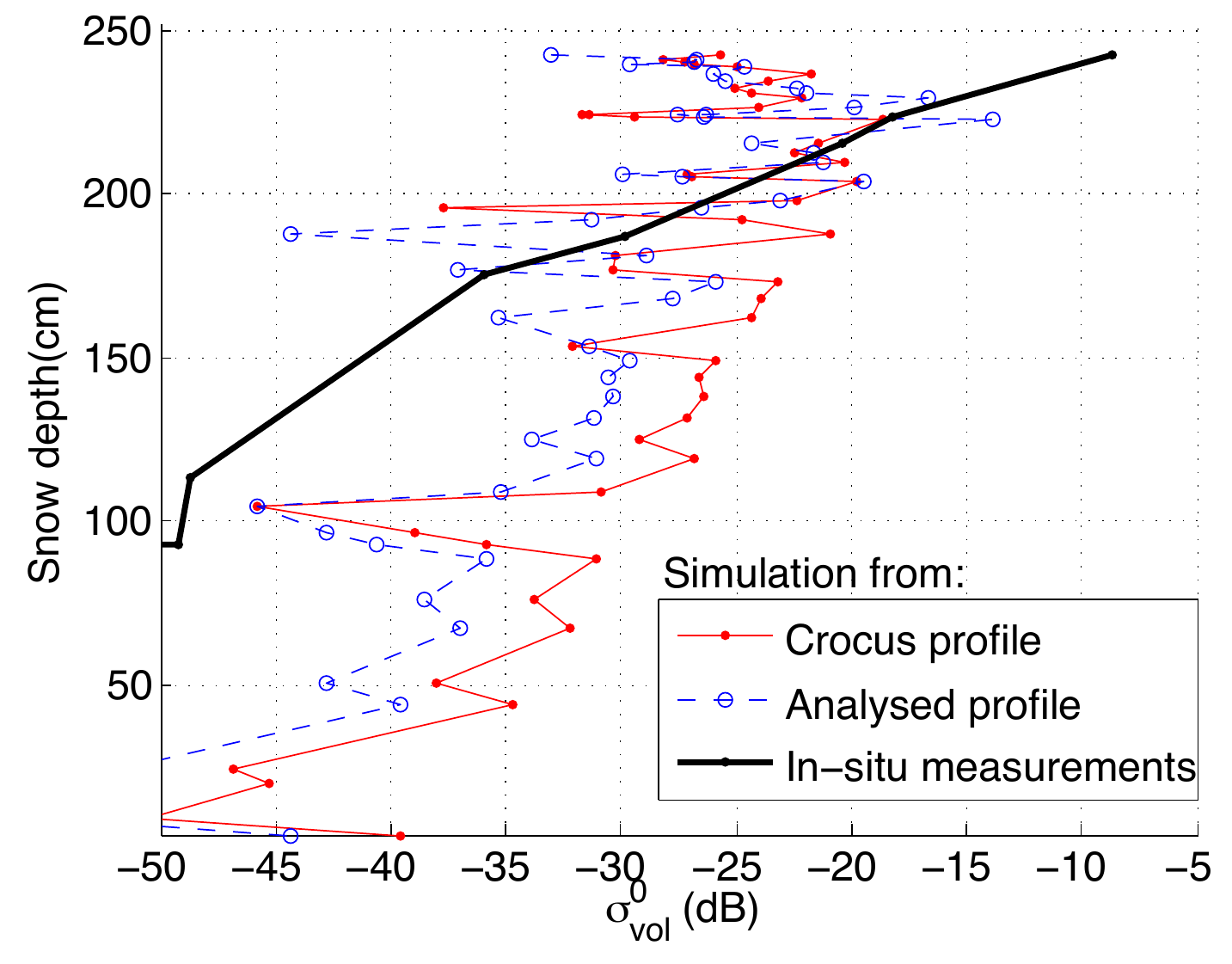}%
\label{fig_0313}}}
\centerline{Date: 13 Mar 2009. Altitude: 2700m.}
\caption{The results of 3D-VAR data analysis method on some Crocus profiles. Each row contains 3 graphs that show the changes made by the analysis algorithm to each layer. Respectively from left to right: the snow optical grain size in mm, density in g/cm$^3$ and the backscattering coefficient of each layer in dB (output of the forward model) using initial Crocus profile and analyzed profile. From up to down are the graphs for the profiles on the altitude of 2700m, of which two in-situ measurements are also available: on 30 January 2009 (plotted on the same graph with Crocus profile on 28 January 2009) and 17 March 2009 (same graph with Crocus profile on 13 March 2009).}
\label{fig:res2}
\end{figure*}

Fig.~\ref{fig:res1} shows the backscattering coefficients obtained over a period of time from three different methods: TerraSAR-X reflectivities, Crocus simulated profiles and simulation of analyzed Crocus profiles. Overall, the differences between the SAR reflectivities and the output of EBM simulation using Crocus initial guess profiles are approximately 2 to 6 dB. The EBM can overestimate the loss of EMW intensity while propagating through the snowpack medium due the assumption that snow particles are of spherical shape. The definition of snow optical grain size, the calculation of the effective permittivity and the phase matrix are also based on the same hypothesis. This assumption does not always hold in the natural environment where snow particles can have various shape and size. It is necessary to develop a more sophisticated method of modeling interaction between EMW and snow particles of different geometry properties.

It can be observed that the gap between the TerraSAR-X backscattering coefficient and the simulation result of assimilated snow parameters is reduced to less than 1 dB. This shows that after the modification made by 3D-VAR, the analysed snowpack stratigraphic profiles give results of simulation closer to the backscattering value observed from radar. Some of the analysed $\sigma_0$ are less closed to the observation than the others (19 February 2009 of 2700m altitude or 2 March 2009 of 3000m altitude). This may due to two reasons. First, the gradient descent using Newton method can converge to local minimum instead of global minimum. With a Crocus profile of 50 snow layers, the algorithm involves of balancing 100 snow parameters in order to find the minimum of the cost function. The probability of having many local minima is significant high. Second, the modeling of the covariance matrix \textbf{B} needed to be further developed to have a more accuracy estimation of guess parameters' errors. Future works need to address these problematic using different kind of optimization method that can reduce the effect of local minimal and developing a better model for the error covariance matrix $\textbf{B}$.

Fig.~\ref{fig:res2} shows the detailed of the modifications of snow stratigraphic profiles done by data analysis process. The input parameters contain the snow optical grain size for Crocus, visually estimated grain size for the in-situ measurements, and the density of each snow layer. It can be observed that the modifications occur mostly on the near-surface layers. This can be due to two reasons:
\begin{itemize}
\item The EMW at higher frequency has lower penetration rate. Depends on the compactness of the snowpack environment, X-band EMW can penetrate from 80 to 120 cm. This means the radar has little sensitivity to the characteristic of snowpack in the deeper layers. The EBM has taken into account this penetration rate through the calculation of attenuation. The deeper EMW penetrate, the higher value of attenuation is accumulated, and therefore the backscattering coefficient of the snow layers decreases exponentially from the surface layer to the ground layer. This can be observed from the graphs on third column of Fig.~\ref{fig:res2}.
\item The error covariance matrix of measurements $\textbf{B}$ is calculated based on the error correlation among layers. This correlation is based strongly on the distance between the layers (\ref{eq:corr}). Large distance between two layers results in low value of correlation. Therefore the modifications of the snow parameters of the near surface layers are considered independent from the deeper layers.
\end{itemize}

\begin{table}[!t]
\caption{\emph{Comparison of bias and root-mean-square deviation (RMSD) between initial Crocus profiles and analysed profiles, with respect to the in-situ measurements}}
\label{tab:bias}
	\begin{center}
		\begin{tabular}{|c|c|c|c|c|}
		\hline
		\hline
		Date & Parameter & Profile & Bias & RMSD\\ \hline
		\multirow{4}{*}{28 Jan 2009} & \multirow{2}{*}{Grain size (mm)} & Crocus & 0.43 & 0.59 \\ \cline{3-5}
									   & & Analysed & 0.45 & 0.57 \\ \cline{2-5}
					     & \multirow{2}{*}{Density (kg/m$^3$)} & Crocus & 110 & 120 \\ \cline{3-5}
									 & & Analysed & 40 & 50 \\ \hline	
		\multirow{4}{*}{13 Mar 2009} & \multirow{2}{*}{Grain size (mm)} & Crocus & 0.49 & 0.64 \\ \cline{3-5}
									   & & Analysed & 0.46 & 0.61 \\ \cline{2-5}
					     & \multirow{2}{*}{Density (kg/m$^3$)} & Crocus & 120 & 130 \\ \cline{3-5}
									 & & Analysed & 50 & 70 \\ \hline \hline
		\end{tabular}
	\end{center}
\end{table}

Two in-situ measurements were carried out on 30 January 2009 (Tab.~\ref{tab:pro_str}) and 17 March 2009. The stratigraphic profiles are plotted on the same graphs as the Crocus profiles of 28 January 2009 and 13 March 2009 (Fig.~\ref{fig:res2}). The total simulated backscattering coefficients of these profiles are also displayed on Fig.~\ref{fig:res1}. We calculate the bias and root-mean-square deviation (RMSD) between the initial (open-loop) Crocus profiles and the in-situ measurements, and compare to the bias and RMSD between the analysed profiles and in-situ measurements. Table~\ref{tab:bias} shows the comparison of these quantities. The results show the bias and RMSD between the analysed snow density and the measurements are much smaller than the initial guess (Crocus) snow density, hence the modifications made by the data analysis tend to approach the in-situ measurement. The analysis however shows little improvement with the modifications on the snow optical grain size, due to two reasons. First, the weight of the snow optical grain size in the covariance error matrix \textbf{B}, as well as in the adjoint model, is bigger than the weight of snow density. Hence it can also be noted that the values of optical grain size are not modified as much as the density (Fig.~\ref{fig:res2}). Second, the grain size used in Crocus is the snow optical radius, where the in-situ measurement uses the visually estimated grain size. Therefore the two quantities are not directly comparable to each other.

\section{Conclusion}

The results of this study show the potential of using data analysis method and the multilayer snowpack backscattering model based on the radiative transfer theory in order to improve the snowpack detailed simulation. The new backscattering model adapted to X-band and higher frequencies enables the calculation of EMW losses in each layer of the snowpack more accurately. Through the use of 3D-VAR data analysis based on the linear tangent and adjoint operator of the forward model, we have the possibility to modify and improve the snowpack profiles calculated by the detailed snowpack model Crocus. The output of this process shows that the discrepancies between the simulated profile and the in situ measurements are smaller after assimilation, and therefore could be further developed and used in real application such as snow cover area monitoring on massif scale or snowpack evolution through a period of time using series of spaceborne SAR image data.

Future studies will be concentrated on developing the assimilation process. The 3D-VAR algorithm needs to be intergrated into Crocus, which means the analysed parameters of each step will be used as the input for the next step of initialization of Crocus. The result will be an intermittent assimilation process where the snow stratigraphic profile generated by Crocus is continuously analysed and adjusted using TerraSAR-X data.

\section*{Acknowledgment}
This work has been funded by GlaRiskAlp, a French-Italian project (2010-2013) on glacial hazards in the Western Alps and M\'et\'eo-France. TerraSAR-X data was provided by German Aerospace Center (DLR). In-situ measurements were carried out by IETR (University of Rennes 1), Gipsa-lab (Grenoble INP) and CNRM-GAME/CEN. The authors would like to thank Gilbert Guyomarc'h, Matthieu Lafaysse and Samuel Morin from CNRM-GAME/CEN in carrying out field measurements and performing the Crocus runs.

\ifCLASSOPTIONcaptionsoff
  \newpage
\fi
\bibliographystyle{IEEEtran}
\bibliography{biblio}
%\end{thebibliography}

%\begin{biography}[{\includegraphics[width=1in,height=1.25in,clip,keepaspectratio]{mshell}}]{Michael Shell}

% \end{IEEEbiography}


%\newpage



%\vfill


\end{document}